\documentstyle[aas2pp4]{article}
\def\showfig#1{{#1}}

\def\comment#1{{}}

\begin{document}

\title{Strong Field Gravity and X-Ray Observations of
4U~1820--30} 

\authoremail{pkaaret@cfa.harvard.edu}

\author{P. Kaaret\altaffilmark{1}, S.
Piraino\altaffilmark{1,2,3}, P.F. Bloser\altaffilmark{1}, E.
C. Ford\altaffilmark{4}, J.E. Grindlay\altaffilmark{1}, A.
Santangelo\altaffilmark{2}, A.P. Smale\altaffilmark{5}, and W.
Zhang\altaffilmark{5}}

\altaffiltext{1}{Harvard-Smithsonian Center for Astrophysics,
60 Garden St., Cambridge, MA 02138, USA}

\altaffiltext{2}{IFCAI/CNR, Via Ugo La Malfa 153, 90146,
Palermo, Italy,}

\altaffiltext{3}{Dipartimento di Scienze Fisiche ed
Astronomich, via Archirafi 36, 90123 Palermo Italy}

\altaffiltext{4}{Astronomical Institute ``Anton Pannekoek,''
Center for High Energy Astrophysics, University of Amsterdam,
Kruislaan 403, 1098 SJ Amsterdam, The Netherlands}

\altaffiltext{5}{NASA/Goddard Space Flight Center, Greenbelt,
MD 20771, USA}

\begin{abstract}

The behavior of quasi-periodic oscillations (QPOs) at
frequencies near 1~kHz in the x-ray emission from the neutron
star x-ray binary 4U~1820-30 has been interpreted as evidence
for the existence of the marginally stable orbit, a key
prediction of strong-field general relativity.  The signature
of the marginally stable orbit is a saturation in QPO
frequency, assumed to track inner disk radius, versus mass
accretion rate.  Previous studies of 4U~1820-30 have used
x-ray count rate as an indicator of mass accretion rate.
However, x-ray count rate is known to not correlate robustly
with mass accretion rate or QPO frequency in other sources. 
Here, we examine the QPO frequency dependence on two other
indicators of mass accretion rate: energy flux and x-ray
spectral shape.  Using either of these indicators, we find
that the QPO frequency saturates at high mass accretion
rates.  We interpret this as strong evidence for the
existence of the marginally stable orbit.

\end{abstract}


\section{Introduction}

For many nuclear equations of state, the surface of a neutron
star is expected to lie within the marginally stable orbit -
the radius within which no stable circular orbits exist within
general relativity (Klu\'{z}niak \& Wagoner 1985).  Thus,
under certain conditions, the marginally stable orbit should
be dynamically important in determining the accretion flow
onto an accreting neutron star.  The millisecond quasiperiodic
oscillations (kHz QPOs) discovered (Strohmayer et al.\ 1996;
van der Klis et al.\ 1996) with the {\it Rossi X-Ray Timing
Explorer} (RXTE) (Bradt, Rothschild, \& Swank 1993) in the
X-ray emission of neutron star x-ray binaries are likely
produced by motion in the inner accretion disk with the QPO
frequency related to the inner disk radius.  Recently, it has
been suggested that the behavior of the kHz QPOs for certain
systems provides evidence for the existence of marginally
stable orbit (Kaaret, Ford, \& Chen 1997; Zhang, Strohmayer,
\& Swank 1997).  The best evidence has come from the binary
4U~1820-30 (Zhang et al.\ 1998a).  

Two simultaneous kHz QPOs are detected from 4U~1820-30 (Smale,
Zhang, \& White 1997; Zhang et al.\ 1998a).  Both the upper
and lower QPO peak vary in frequency, but maintain an
approximately constant frequency separation.  Zhang et al.\
(1998a) found that below a certain count rate, the frequency
centroids of both the upper and lower QPO peaks are correlated
with count rate, while above that count rate, the QPO
frequencies are roughly constant and are independent of count
rate.  Saturation of QPO frequency at high mass accretion
rates had been suggested as a signature of the marginally
stable orbit (Miller, Lamb, \& Psaltis 1998; Kaaret et al.\
1997).  A key question is whether the x-ray count rate is a
good indicator of the mass accretion rate and whether it is
robustly correlated with the QPO frequency.  In general, x-ray
count rate is not a good indicator of mass accretion rate
(Hasinger \& van der Klis 1989).

Here, we present new observations of 4U~1820-30 made
simultaneously with RXTE and BeppoSAX and a new analysis of
archival RXTE observations.  The new data show that the QPO
frequency is not robustly correlated with x-ray count rate in
4U~1820-30.  Thus a better indicator of mass accretion rate is
required.  We use energy flux and x-ray spectral shape as
indicators of mass accretion rate. Below, we first discuss the
observations and analysis. We then present correlations
between the QPO frequency and various indicators of mass
accretion rate.  We conclude with a discussion of the results.

\section{Observations and Analysis}

We performed two joint BeppoSAX/RXTE observations of
4U~1820-30, on 1998 April 17-18 and 1998 September 19-20, for
a total of 100~ks of on-source observing time in each of
BeppoSAX and RXTE.  We also reanalyzed the RXTE observations
described in Smale et al.\ (1997) and Zhang et al.\ (1998a).

For the timing analysis, $122 \, \mu$s time resolution event
data from the RXTE Proportional Counter Array (PCA) were used.
To search for fast QPOs, we performed Fourier transforms on
2~s segments of data with no energy selection and summed the
2~s power spectra within each observation interval.  The total
power spectrum for each interval was searched for QPO peaks
above 200~Hz by fitting a function, consisting of a Lorentzian
plus a constant, for each trial frequency.  Only QPO peaks
with a chance probability of occurrence of less than 1\% as
determined from an F-test were retained.

\begin{figure}[tb] \showfig{\epsscale{0.9}
\plotone{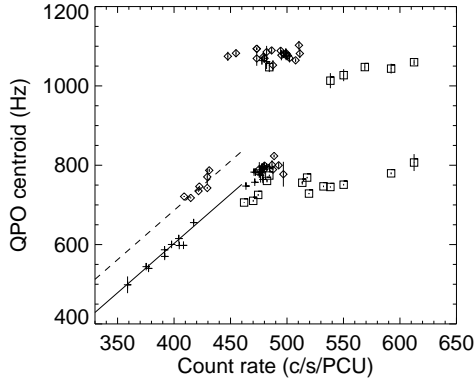}} \caption{QPO centroid frequency versus
x-ray count rate in the RXTE PCA (full band PCA with
background subtracted).  The diamonds indicate our new data
from 1998.  The squares are observations from 1997 and the
crosses from 1996.  The solid line is a best fit to the 1996
data in the count rate range $340-440 \rm \, c \, s^{-1} \,
PCU^{-1}$.  The dashed line is offset by 83.5~Hz.  The offset
of the 1996 versus 1998 data show that lower QPO frequency is
not robustly correlated with x-ray count rate.}
\label{qpo_rate} \end{figure}

\begin{figure}[tb] \showfig{\epsscale{0.9}
\plotone{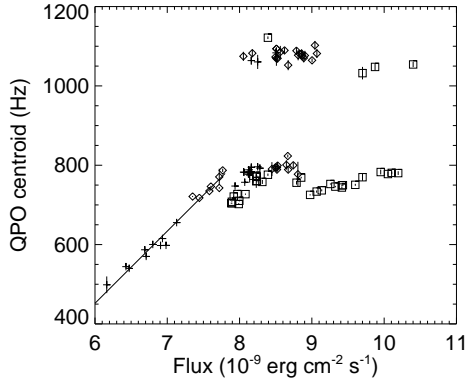}} \caption{QPO centroid frequency versus
unabsorbed energy flux in the 2--25~keV band calculated from
the PCA spectra.  The plot symbols indicate the epoch of the
observations as in Fig.~\ref{qpo_rate}.  The line is a fit to
the data in the flux range $6-7.8 \times 10^{-9} \rm \, erg \,
cm^{-2} \, s^{-1}$.  The lower QPO frequency appears robustly
correlated with energy flux at low fluxes and saturates at
high fluxes.} \label{qpo_flux} \end{figure}

For the RXTE-only analysis, presented in Figs.~\ref{qpo_rate},
\ref{qpo_flux}, \ref{qpo_hc}, and \ref{hc_flux2}, we divided
the data into continuous segments with lengths near 1000~s. 
In some cases several 1000~s of data were combined to allow
detection of weaker QPOs.  For each interval, we calculated
PCA count rates in various energy bands using Proportional
Counter Units (PCUs) 0, 1, and 2 as these were on in all
observations.  As the PCA response changed gradually over the
2 year span of these observations, we used fixed energy bands
and interpolated the count rates within fractional channel
boundaries.  The x-ray colors in Figs.~\ref{qpo_hc} and
\ref{hc_flux2} were calculated from these rates.  We also
found 129 channel PCA spectra for PCUs 0 and 1, which were
used to calculate the unabsorbed energy fluxes for the
2--25~keV band presented in Figs.~\ref{qpo_flux} and
\ref{hc_flux2}.  Detailed analysis of RXTE spectral variations
and correlations with the QPOs will be given in Bloser et al.\
(1999).

\begin{figure}[tb] \showfig{\epsscale{0.9}
\plotone{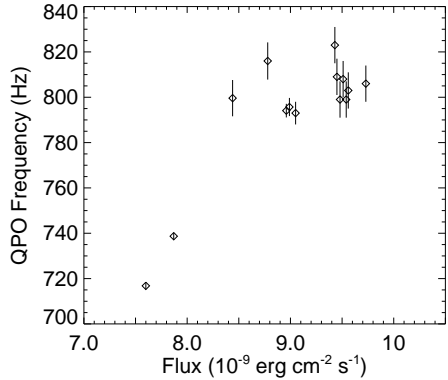}} \caption{Correlation of QPO
centroid frequency from RXTE observations with the unabsorbed
energy flux in the 0.3--40~keV band calculated from
simultaneous BeppoSAX observations.  Only the lower frequency
QPO is shown.  The QPO frequency  appears independent of flux
above $8.3 \times 10^{-9} \rm \, erg \, cm^{-2} \, s^{-1}$.}
\label{qpo_saxflux} \end{figure}

For the simultaneous BeppoSAX/RXTE analysis we selected the
longest continuous segments, in order to maximize the
statistics in each spectrum.  There are 26 simultaneous
segments, for a total of 62~ks.  We searched for QPOs in the
PCA data and performed spectral analysis of the BeppoSAX data
for each segment.  The spectral analysis is described in
Piraino et al.\ (1999).  We chose to use a Comptonization
model (Sunyaev \& Titarchuk 1980) with an added single
temperature standard blackbody component and including
interstellar absorption to parameterize the spectra.  For this
model, the temperature, $T_{c} = 2.83 \pm 0.08 \rm \, keV$,
and optical depth, $\tau = 13.7 \pm 0.5$, of the Comptonizing
electron cloud, the flux of the blackbody component, and the
absorption column density, $N_H = (2.8 \pm 0.3) \times 10^{21}
\rm \, cm^{-2}$, are constant within errors for all the joint
BeppoSAX/RXTE observation intervals (the uncertainties quoted
are the typical 90\% confidence uncertainty for individual
spectra).  The values found for $\tau$ and $T_{c}$ are similar
to those observed previously (Christian \& Swank 1997), and
the parameters of the blackbody are similar to those found in
an ASCA observation of 4U~1820-30 in a low intensity state
(Smale et al.\ 1994).  The blackbody temperature, $T_{bb}$,
and the total flux vary with QPO frequency as reported in
Piraino et al.\ (1999).  Using this model, we calculated the
unabsorbed energy flux in the 0.3-40~keV band  presented in
Fig.~\ref{qpo_saxflux}.

\section{Correlation of Spectral and Timing Properties}

Zhang et al.\ (1998a) showed that the QPO frequency in
4U~1820-30 is correlated with x-ray count rate below a certain
critical count rate and that the QPO frequency saturates above
that count rate.  This has been interpreted as evidence for
the existence of the marginally stable orbit.  However, a
critical question, whether the x-ray count rate is a reliable
estimator of the mass accretion rate through the disk (Kaaret
et al.\ 1998), needs to be addressed for 4U~1820-30.  

Zhang et al.\ (1998a) includes observations from 1996 and
1997, but the data which show a correlation between x-ray
count rate and the lower QPO frequency come from observations
which occurred within one 8~hour period in 1996.  As the QPO
frequency versus count rate relation is known to change on
time scales longer than days (Ford et al. 1997a; Yu et al.\
1997; Zhang et al.\ 1998b; Mendez et al.\ 1999), it is
important to check this relation with additional
observations.  Figure~\ref{qpo_rate} shows the QPO frequency
plotted versus PCA count rate for RXTE observations spanning 2
years.  This plot shows both the upper frequency (above
1000~Hz) and lower frequency (below 850~Hz) QPOs.  The varying
upper frequency peaks reported in Zhang et al.\ (1998a) appear
in the data, but have F-test values in the range 1--4\% and,
thus, are not included here.  Both the upper and lower QPO
frequencies saturate at high count rates.  Below a count rate
of $440 \rm \, c \, s^{-1} \, PCU^{-1}$, the lower QPO
frequency appears correlated with count rate.  However, our
new data show a shift of the lower QPO frequency versus count
rate relation by $83.5 \pm 9.0 \rm \, Hz$ relative to the 1996
data.  This shift indicates that the x-ray count rate versus
QPO frequency correlation is not robust in 4U~1820-30. Thus,
other indicators of mass accretion rate are required.

Fig.~\ref{qpo_saxflux} shows the relation between the lower
QPO frequency and the broad-band (0.3--40~keV) unabsorbed
energy flux calculated from the BeppoSAX data. The QPO
frequency appears uncorrelated with the unabsorbed energy flux
at high fluxes.  The same behavior is seen for the absorbed
flux.  As the broad-band spectral coverage of BeppoSAX allows
a reasonably reliable estimate of the total unabsorbed flux,
this indicates that the break in the QPO frequency versus
count rate relation is not simply due to a spectral change.  

We also examined the QPO frequency versus flux relation with
the full RXTE data set, see Fig.~\ref{qpo_flux}.  Due to the
PCA's lack of spectral coverage below 2~keV, reliable
estimates of the broad-band energy flux are not available, so
we chose to calculate the unabsorbed energy flux in the
2.0--25~keV band from the PCA data.  The lower QPO frequency
appears correlated with flux in observations spanning 2
years.  This is in contrast to the lack of robust correlation
between QPO frequency and flux in other sources (Ford et al.
1997b; Zhang et al. 1998b).  In 4U~1820-30, the QPO frequency
saturates at high fluxes, consistent with the signature of the
marginally stable orbit.

\begin{figure}[tb] \showfig{\epsscale{0.9}
\plotone{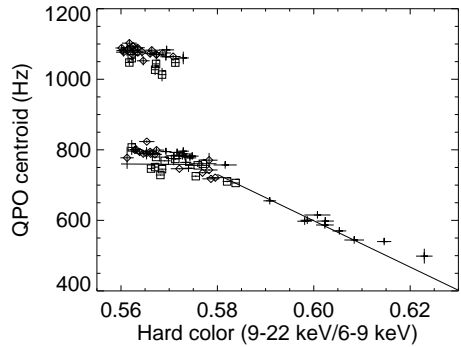}} \caption{QPO centroid frequency versus
an x-ray hard color $C$ defined as the ratio of counts in the
9--22~keV band to that in the 6--9~keV band calculated from
the PCA data.  Lower color values correspond to higher mass
accretion rates.  The solid lines are linear fits of the lower
QPO frequency versus hard color for $C > 0.58$ and $C <
0.575$.  The plot symbols indicate the epoch of the
observations as in Fig.~\ref{qpo_rate}.  The lower QPO
frequency appears robustly correlated with x-ray color for
high color values, and saturates for low color values.}
\label{qpo_hc} \end{figure}

Kaaret et al.\ (1998) showed that the spectral shape at high
energies, above 5~keV, correlates well with QPO frequency for
the atoll sources 4U~0614+091 and 4U~1608-52.  Spectral shape
is also generally accepted as a good indicator of mass
accretion rate in neutron-star low mass x-ray binaries based
on studies of their spectra and timing noise (e.g. Hasinger \&
van der Klis 1989; Schulz, Hasinger, \& Trumper 1989).   The
ratio of counts in two energy bands, an x-ray color, can be
taken as estimator of spectral shape.  In Fig.~\ref{qpo_hc},
we show the relation between QPO frequency versus a hard x-ray
color, $C$, defined as the ratio of counts in the 9--22~keV
band to that in the 6--9~keV band and calculated from the PCA
data.  The lower QPO frequency is well correlated with x-ray
color when the spectrum is hard.  There appears to be a break
to constant QPO frequency when the spectrum becomes softer
than a certain critical value.  

To test whether the break in the QPO frequency versus hardness
relation is significant, we did linear fits of the lower QPO
frequency versus hard color for $C > 0.58$ and $C < 0.575$. 
For $C > 0.58$, the linear correlation coefficient is
$-0.984$, corresponding to  a chance probability of occurrence
of $1.4 \times 10^{-9}$, and the best-fit slope is $-6560 \pm
350 \rm \, Hz/color \, unit$. While for $C < 0.575$, the
correlation coefficient is $-0.012$, which implies that an
uncorrelated data set would produce a correlation coefficient
of this magnitude or larger with a probability of 0.97, and
the slope is $-97 \pm 1310 \rm \, Hz/color \, unit$.  We also
fitted the entire data set for the lower frequency peak to two
models: a line and a broken line with saturation at a maximum
frequency.  The $\chi^2$ improved with addition of the extra
parameter for the frequency saturation, $\Delta \chi^2 /
\chi^2_{\nu} = 28.74$.  The addition of the extra parameter is
justified at a confidence level of $2 \times 10^{-6}$.  Also,
the slope found in the broken-line fit agrees well with the
slope found from the linear fit for $C > 0.58$.  Thus, the
evidence for the saturation in QPO frequency at low $C$, i.e.
high mass accretion rates, is highly significant when
evaluated using the linear correlation coefficients, the
slopes of the linear fits in the two color ranges, or the
decrease in $\chi^2$ with the addition of saturation to the
linear model.

\begin{figure}[tb] \showfig{\epsscale{0.9}
\plotone{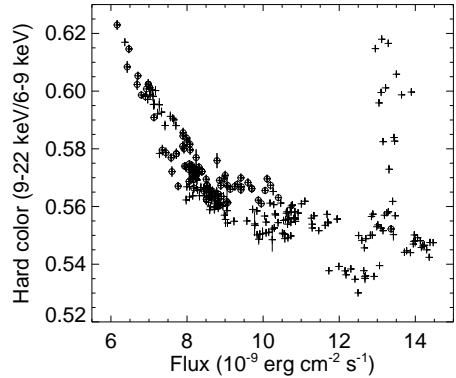}} \caption{X-ray hard color versus energy
flux, both calculated from the PCA data.  The circles indicate
intervals during which kHz QPOs were detected.  The hard color
versus flux relation has no features at the point where the
QPO frequency saturation occurs. The QPO at very high flux has
a centroid of $290.7 \pm 1.2 \rm \, Hz$ and a width of 13~Hz
and may indicate a direct detection of the neutron star spin;
however, the detection is not of high significance.}
\label{hc_flux2} \end{figure}

In most of the atoll kHz QPO sources studied, QPOs are
detected only down to some minimum hard color (e.g. Mendez
1999).  The QPOs generally disappear near, or just beyond, a
turning point in the color-flux or soft color-hard color
diagram and near the cutoff the QPO frequency varies over a
wide range, 200-300~Hz.  In 4U~1820-30, the QPOs disappear at
a hard color near 0.56 where there is a break in the
color-flux relation, see Fig.~\ref{hc_flux2}.  This is similar
to the behavior seen in other sources, although the range of
the QPO frequency variation near the QPO disappearance is
smaller in 4U~1820-30.

The saturation of QPO frequency in 4U~1820-30 occurs at a
hard color near 0.58, significantly higher than the hard
color where the QPOs disappear.  The saturation occurs at a
position in the color-flux (or color-color) diagram where
other kHz QPO sources show a good correlation between QPO
frequency and hard color, and is thus markedly distinct from
the behavior seen in the other kHz QPO sources.  The fact
that the QPO frequency saturates while the hard color
continues to increase suggests that the break in QPO
frequency seen in 4U~1820-30 is distinct from the QPO
behaviors seen in other sources.  Because QPO frequency is
well correlated with hard color in most kHz QPO sources and
because x-ray color is thought to be a good indicator of mass
accretion rate, the break in the QPO frequency versus x-ray
hard color relation strengthens the interpretation of the
transition to constant QPO frequency in 4U~1820--30 as
evidence for the marginally stable orbit.

\section{Discussion}

Interpretation of the kHz QPO data as evidence for the
detection of the marginally stable orbit requires that the 
mechanism producing the kHz QPOs have a frequency which
increases monotonically as the inner disk radius decreases. 
Keplerian orbital motion, sub-Keplerian orbital motion,
perhaps due to radiation pressure, or coherent phenomena, such
as sound waves or disk oscillations modes, would all give the
requisite dependence of QPO frequency on inner disk radius. 
In any of these cases, the sharp break in the mass accretion
rate dependence of the QPO frequency in 4U~1820-30 requires a
correspondingly sharp break in the dependence of inner disk
radius on mass accretion rate.  

The transition in QPO behavior versus x-ray hard color shown
in Fig.~\ref{qpo_hc} occurs over an interval only slightly
larger than the typical accuracy of the hard color
measurements.  Thus, the break must occur over a narrow range
in mass accretion rate.  A change in disk structure could
produce a change in the dependence of QPO frequency on mass
accretion rate (Zhang et al. 1998b).  However, it is unlikely
that a change in disk structure would produce a change to a
fixed inner disk radius that then remains constant as the mass
accretion rate continues to increase.  It is possible that the
break is due to some special orbital radius other than the
marginally stable orbit.  Ghosh (1998) recently discovered an
accretion disk instability that occurs in a fixed annulus
covering radii of $14-19 \, GM/c^2$ for a neutron star mass
$M$.  However, the instability is important only for gas
pressure dominated disks, and is unlikely to be important for
4U~1820-30 as the mass accretion rate is of order 0.1 of the
Eddington rate, and thus the disk is likely radiation pressure
dominated.

An alternative explanation, that the accretion disk is
terminated at the neutron star surface, is rejected because
the high coherence, $\nu / \Delta \nu \sim 30$, of the QPOs
requires a life time for the phenomena producing the QPOs of
at least 30 orbital or rotational cycles.  Any spatially
localized or coherent phenomena at the inner edge of the disk
would be rapidly disrupted by the viscous stress and magnetic
fields at the neutron star surface if the disk is terminated
at the stellar surface (Miller et al.\ 1998).  Thus, the
observed coherence could not be maintained.  Another
alternative explanation, that the mass accretion rate
independent QPO frequency is the spin frequency of the neutron
star, is rejected because, based on the persistent emission
and x-ray burst QPOs detected from other atoll sources, the
spin frequency of 4U~1820-30 is most likely within a few 10~Hz
of the QPO difference frequency of 270~Hz.

The QPO frequency above the critical mass accretion rate is
not constant, but appears to vary over a fractional range (for
the upper QPO) of approximately 5\%.  Analysis of accretion
disk flow across the marginally stable orbit (Muchotrzeb 1983;
Muchotrzeb-Czerny 1986) shows that if the inner disk radius is
driven near the marginally stable orbit, then the inner disk
radius varies over a range consistent with the observed QPO
frequency variations (Kaaret et al.\ 1997).

We conclude that the observations of millisecond QPOs in the
x-ray emission from 4U~1820-30 provide the first strong
experimental evidence for the existence of the marginally
stable orbit.

\acknowledgments  

We gratefully acknowledge the efforts of Evan Smith (RXTE
mission planner) and the BeppoSAX mission planning team in
coordinating the observations.  We thank the referee, Hale
Bradt, for many useful comments.  PK and SP acknowledge
support from NASA grants NAG5-7405 and NAG5-7334.




\begin{references}

\reference{} Bloser, P.F. et al.\ 1999, in preparation.

\reference{} Bradt, H.V., Rothschild, R.E., \& Swank, J.H.
1993, AAS, 97, 355

\reference{} Christian, D.J. \& Swank, J.,H. 1997 ApJS, 109,
177

\reference{} Ford, E.C. et al. 1997a, \apjl, 475, L123

\reference{} Ford, E.C. et al. 1997b, \apjl, 486, L47

\reference{} Ghosh, P. 1998, \apjl, 506, L109

\reference{} Hasinger, G. \& van der Klis, M. 1989, \aap, 225,
79

\reference{} Kaaret, P., Ford, E. C. \& Chen, K. 1997, ApJ,
480, L27

\reference{kaaret97b} Kaaret, P. \& Ford, E.C. 1997, Science,
276, 1386

\reference{} Kaaret, P., Yu, W., Ford, E.C., \& Zhang, S.N.
1998, ApJ, 497, L93

\reference{} Kluzniak, W. \& Wagoner, R.V.  1985, \apj, 297,
548

\reference{} Mendez, M. 1999, Proceedings of the 19th Texas
Symposium in Paris, to appear

\reference{} Mendez, M., et~al. 1999, \apjl, 511, L49

\reference{} Miller, M. C., Lamb, F. K., \& Psaltis, D. 1998,
ApJ, 508, 791

\reference{} Muchotrzeb, B. 1983, Acta Astronomica, 33, 79

\reference{} Muchotrzeb-Czerny, B. 1986, Acta Astronomica, 36,
1

\reference{} Piraino, S. et al. 1999, Proceedings of the 19th
Texas Symposium in Paris, to appear

\reference{} Schulz, N.S., Hasinger, \& G., Trumper, J. 1989,
\aap, 225, 48

\reference{} Sunyaev $\&$ Titarchuk 1980, A$\&$A, 86, 121

\reference{} Smale, A.P., et~al., 1994, BAAS, 184

\reference{} Smale, A.P., Zhang, W., \& White, N. E.,  1997,
ApJ, 483, L119

\reference{} Strohmayer, T.E. et al. 1996, \apjl, 469, L9

\reference{} van der Klis, M. et al. 1996, \apjl, 469, L1

\reference{} Yu, W. et al. 1997, \apjl, 490, L153

\reference{} Zhang, W., Strohmayer, T.E., \& Swank, J.H. 1997,
\apjl, 482, L167

\reference{} Zhang, W. et~al. 1998a, \apjl, 500, L171

\reference{} Zhang, W. et al. 1998b, \apjl, 495, L9


\end{references}
\end{document}